\documentclass[letterpaper,12pt]{article}
\usepackage{color,amssymb,enumerate,float,amsmath}

\usepackage{ifpdf}
\ifpdf
	\usepackage[pdftex]{graphicx}
	\usepackage[pdftex,unicode,implicit]{hyperref}

	\hypersetup{
  	pdftitle     = {}, 
  	pdfkeywords  = {},
  	pdfauthor    = {},
  	pdfcreator   = {pdf\LaTeXe\ with package \flqq hyperref\frqq},
  	pdfproducer  = {pdf\LaTeXe\ with package \flqq hyperref\frqq},
  	pdfpagemode  = UseNone,  
  	pdffitwindow = true,  
  	unicode      = true,
  	plainpages   = true,
  	colorlinks   = true,  
  	citecolor    = black,  
  	urlcolor     = blue, 
  	linkcolor    = black
	}

\else

  \usepackage[dvips]{graphicx}

\fi 

\makeatletter
\@addtoreset{equation}{section}
\makeatother

\begin{document}
	
\thispagestyle{empty}

\begin{center}
{\bf \LARGE Quantization of the non-projectable $2+1\rm{D}$ Ho\v{r}ava theory: The second-class constraints}
\vspace*{15mm}

{\large Jorge Bellor\'{\i}n}$^{1}$
{\large and Byron Droguett}$^{2}$
\vspace{3ex}

{\it Department of Physics, Universidad de Antofagasta, 1240000 Antofagasta, Chile.}
\vspace{3ex}

$^1${\tt jbellori@gmail.com,} \hspace{1em}
$^2${\tt byron.droguett@ua.cl}

\vspace*{15mm}
{\bf Abstract}
\begin{quotation}{\small\noindent
We present the quantization of the $2+1$ dimensional nonprojectable Ho\v rava theory. The central point of the approach is that this is a theory with second-class constraints, hence the quantization procedure must take account of them. We consider all the terms in the Lagrangian that are compatible with the foliation-preserving-diffeomorphisms symmetry, up to the $z=2$ order which is the minimal order indicated by power-counting renormalizability. The measure of the path integral must be adapted to the second-class constraints, and this has consequences in the quantum dynamics of the theory. Since this measure is defined in terms of Poisson brackets between the second-class constraints, we develop all the Hamiltonian formulation of the theory with the full Lagrangian. We found that the propagator of the lapse function (and the one of the metric) acquires a totally regular form. The quantization requires the incorporation of a Lagrange multiplier for a second-class constraint and fermionic ghosts associated to the measure of the second-class constraints. These auxiliary variables have still nonregular propagators.
}
\end{quotation}
\end{center}

\thispagestyle{empty}

\newpage
\section{Introduction}
Ho\v rava theory \cite{Horava:2009uw,Horava:2008ih} is a geometrical field theory that may be used to study quantum gravity since it is power-counting renormalizable and unitary. As a field theory, it has some open questions that deserve deep analysis. In particular, the consistent quantization of its nonprojectable version must be focused. This is a rather nondirect program since the nonprojectable theory has second-class constraints, hence any scheme of quantization must take account of them. Once a consistent framework for such a quantized theory has been established, an important application for it is to prove (or disprove) its renormalizability. On the contrary, the projectable version has not second-class constraints, hence its quantization can be achieved with the standard techniques of gauge field theories. Indeed, the renormalizabilty of the projectable version has been proven \cite{Barvinsky:2015kil}. Moreover, it is a theory with asymptotic freedom in 2+1 dimensions \cite{Barvinsky:2017kob}. Quantum corrections to the $2+1$ projectable theory has been studied in Ref.~\cite{Griffin:2017wvh}.

The geometrical framework introduced in the Ho\v rava theory is to represent the gravitating space as a foliation of spacelike hypersurfaces along a given direction of time. The foliation is considered as absolute, that is, it cannot be changed by a symmetry transformation. The theory is formulated in terms of the Arnowitt-Deser-Misner (ADM) variables: the spatial metric $g_{ij}$, the lapse function $N$ and the shift vector $N_i$, which are natural for such a foliation \cite{Arnowitt:1962hi}. The gauge symmetry of the theory is given by the group of all the foliation-preserving diffeomorphisms (FDiff) acting on these variables. We emphasize that the presence of this gauge symmetry does not guarantee that the standard quantization procedures for gauge theories --Faddeev-Popov \cite{Faddeev:1967fc}, BRST, background-field method, etc., are sufficient to perform its quantization, due to the presence of the second-class constraints that are not associated to gauge symmetries. Hence, besides the quantization of the gauge sector, one must find a way to incorporate the second class constraints as restrictions on the phase space, making a consistent quantization.

The FDiff gauge symmetry leads to a theory that includes higher order terms in spatial derivatives. As a consequence, it is expected that the renormalizability of the theory is improved with respect to general relativity, since the behavior of propagators is improved in the ultraviolet. Simultaneously, unitarity can be safe since no higher time derivatives are generated. The theory has two versions, the projectable version where the lapse function $N$ is a function only of time and the nonprojectable version where $N$ may depend on the time and the space. The projectable condition is preserved by the FDiff summetry group, hence the projectable case constitutes an independent formulation. The nonprojectable  version has field equations closer to the Einstein equations, and its $3+1$ formulation has more chance of surviving the observational tests than the projectable case. In the nonprojectable version, a fundamental extension of the Lagrangian was proposed in Ref.~\cite{Blas:2009qj} by including terms depending on the FDiff-covariant vector $a_{i}=\partial_{i}\ln N$.

Computations in quatum gravity are typically difficult. In the case of the nonprojectable $3+1$ Ho\v{r}ava gravity, the Lagrangian includes a number of the order of $10^2$ different terms that are compatible with the FDiff gauge symmetry. However, when the dimensionality of the space is reduced to $2+1$, the number of independent terms in the Lagrangian reduces drastically and the theory is still interesting for doing quantum gravity. Indeed, an outstanding feature of the $2+1$ Ho\v{r}ava theory is that it propagates a physical, scalar, degree of freedom, unlike $2+1$ general relativity which is a topological theory. Thus, $2+1$ Ho\v{r}ava theory is a three-dimensional model with a particle carrying the local gravitational interaction, hence, in principle,  perturbative quantization based on Feynman propagators and Feynman diagrams can be used. An exception occurs when the theory is formulated at the critical point, in which case a physical mode is suppressed. This critical point, whose definition depends on the dimensionality of the theory,  has been called the kinetic-conformal point \cite{Bellorin:2016wsl}. In the $2+1$ dimensional case the theory becomes topological at the critical point, since the only propagating mode disappears. This case deserves a separate study, here we only consider the noncritical formulation.

Our objective in this paper is to perform a detailed analysis of the quantization of the $2+1$ nonprojectable Ho\v{r}ava theory. As we have commented, an essential feature of the nonprojectable theory is the presence of second-class constraints. This fact forces us to consider the quantization in rather different approaches to ones used in general relativity, projectable Ho\v{r}ava theory and gauge theories in general with only first-class constraints. As it is well known, there are two main routes to manage theories with second-class constraints, namely, the path integral quantization with the appropriate measure for the second-class constraints and the Dirac brackets in the operator formalism. We focus on the path integral quantization, since it is more adaptable to a gravitational field theory as the Ho\v rava theory. Since the measure corresponding to the second-class constraints is defined in terms of the canonical variables and their Poisson brackets \cite{Senjanovic:1976br}, our approach is based on the Hamiltonian formulation of the theory.

We consider the full $2+1$ theory. Hence we consider in the Lagrangian all the inequivalent terms that are compatible with the FDiff symmetry, up to the minimal order in spatial derivatives required by power-counting renormalizability, which is $z=2$ in the $2+1$ theory \cite{Horava:2009uw}. This yields a Lagrangian with terms of second and fourth order in spatial derivatives. We combine the perturbative approach, where the constraints can be solved and one can obtain the propagator of the theory, with formal nonperturbative and nonreduced approaches.

We comment that, in spite of being a nontopological theory, the $2+1$ Ho\v{r}ava theory still shares some features with $2+1$ general relativity. A fundamental issue with consequences in the local quantization is the definition of asymptotic flatness, which in particular is relevant for the Hamiltonian formulation of the theory. In $2+1$ general relativity the definition of asymptotic flatness is not based on having a fixed metric at infinity, unlike the $3+1$ case. The definition lies on the existence of the exact solution corresponding to the gravitational field of a massive point particle. This solution is a locally flat cone with a deficit angle that depends on the mass of the particle \cite{Deser}. An asymptotically flat configuration is then a configuration that approaches this solution for large enough distances. As a consequence, the dominant mode in the expansion is not fixed functionally, as we have commented. In a previous paper \cite{Bellorin:2019zho}, we studied the analogous situation in the $2+1$ nonprojectable Ho\v{r}ava theory, finding that the same solution for the massive point particle is valid in the Ho\v{r}ava theory. Thus, we proposed the same definition of asymptotic flatness as in $2+1$ general relativity \cite{Ashtekar:1993ds} for the three-dimensional Ho\v{r}ava theory. More consequences on the value of the energy and the role of the higher order terms were considered in that reference.


\section{The nonprojectable Ho\v{r}ava theory in 2 spatial dimensions}
The starting point is the definition of a foliation formed by two-dimensional spatial slices, the foliation being defined along a direction of time. This setting is considered as absolute, it cannot be changed by a symmetry transformation. Thus, the underlying gauge symmetry group is given by the diffeomorphisms that preserve the foliation, FDiff. Under a FDiff transformation, the coordinates $(t,\vec{x})$  transform as  
\begin{equation}
\delta t=f(t),\qquad \delta x^{i}=\zeta^{i}(t,\vec{x}).
\end{equation}
The Ho\v{r}ava theory is formulated in terms of the Arnowitt-Deser-Misner (ADM) variables, which are the spatial metric $g_{ij}$, the lapse function $N$ and the shift function $N_{i}$. Under a FDiff transformation, the ADM variables transform as
\begin{eqnarray}
\delta N &=& 
\zeta^{k} \partial_{k} N + f \dot{N} + \dot{f}N \,, 
\label{deltaN}
\\ 
\delta N_{i} &=& 
\zeta^{k} \partial_{k} N_{i} + N_{k} \partial_{i} \zeta^{k} + \dot{\zeta}^{j} g_{ij} + f \dot{N}_{i} + \dot{f}N_{i} \,,
\\
\delta g_{ij} &=& \zeta^{k} \partial_{k} g_{ij} + 2g_{k(i}\partial_{j)} \zeta^{k} + f \dot{g}_{ij} \,.
\end{eqnarray}
With the FDiff gauge symmetry one may define the Ho\v{r}ava theory in the projectable and the nonprojectable formulations. In the projectable version the lapse function is a function only on time, $N(t)$. This condition is preserved by the FDiff symmetry, as it can be directly deduced from (\ref{deltaN}). We will focus in the nonprojectable version and the minimal degree of anisotropy needed to ensure the power-counting renormalization of the Ho\v rava theory, which is $z = \mbox{spatial dimensions} = 2$ \cite{Horava:2009uw}. 

The action of the nonprojectable Ho\v rava theory in $2+1$ dimensions is 
\begin{equation}
S =
\int dt d^{2}x \sqrt{g} N 
\left( G^{ijkl} K_{ij} K_{kl} - \mathcal{V} \right) \,,
\label{action}
\end{equation}
 where the extrinsic curvature is defined by
\begin{equation}
\label{k} K_{ij}=\frac{1}{2N}\left(\dot{g}_{ij}-2\nabla_{(i}N_{j)}\right),
\end{equation}
 and the hypermatrix $G^{ijkl}$ is a four-index metric 
\begin{equation}
\label{G} G^{ijkl}=\frac{1}{2}\left(g^{ik}g^{jl}+g^{il}g^{jk}\right) -\lambda g^{ij}g^{kl} \,.
\end{equation}
The covariant derivative $\nabla_i$ and all the standard notation of Riemannian geometry that we use refers to the three-dimensional spatial metric $g_{ij}$. We use the shorthand notation $\nabla_{ijk...} \equiv \nabla_i \nabla_j \nabla_k \cdots$ . The dimensionless parameter $\lambda$ included in the kinetic term plays  a fundamental role in the theory. The matrix $G^{ijkl}$ has inverse if and only if $\lambda \neq 1/2$, a condition that we assume throughout this paper. On the contrary, when this parameter acquires the critical value $\lambda=1/d$, where $d$ is the spatial dimension, the kinetic term acquires a conformal invariance, although the whole theory is not conformally invariant due to the potential that breaks this symmetry. The theory at $\lambda = 1/2$ can become conformally invariant if also the potential is conformal, like a Cotton-square term or the potential studied in \cite{Bellorin:2018blt}. At the critical point $\lambda = 1/2$ the extra scalar mode is eliminated due to the raising of two additional second-class constraints \cite{Bellorin:2013zbp}.

The potential $\mathcal{V}$ must be invariant under FDiff. Hence, it must be formed by invariants written in terms of the spatial metric and the acceleration vector \cite{Blas:2009qj} 
\begin{equation}
a_{k}=\partial_{k}\ln N \,.
\label{aceleracion}
\end{equation}
The full potential in $2+1$ dimensions, considering all the terms up to the $z=2$ degree of anisotropy, is given by \cite{Sotiriou:2011dr}
\begin{eqnarray}
\mathcal{V} &=&
-\beta R-\alpha a^{2}+\alpha_{1}R^{2}+\alpha_{2}a^{4}+\alpha_{3}R a^{2}+\alpha_{4}a^{2}\nabla_{k}a^{k}
\nonumber \\ &&
+\alpha_{5} R\nabla_{k}a^{k} 
+\alpha_{6}\nabla^{l}a^{k}\nabla_{l}a_{k}+ \alpha_{7}(\nabla_{k}a^{k})^{2} \,,
\label{potencial}
\end{eqnarray}
where $\beta,\alpha$, and the $\alpha_{1,\ldots,7}$ are independent coupling constants.

\section{Canonical formulation}
\subsection{Hamiltonian and constraints}
Our main aim is to address the path-integral quantization of the theory. To this end we perform the canonical formulation. The nonreduced phase space is spanned by the conjugate pairs $(g_{ij},\pi^{ij})$ and $(N,P_{N})$. The reduced phase is the subspace where all the constraints are satisfied. The first (primary) constraint that arises in the formulation is the vanishing of the momentum conjugate to the lapse function, 
\begin{equation}
P_{N}=0 \,,
\label{Pn}
\end{equation}
since the Lagrangian in (\ref{action}) does not depend on the time derivative of $N$. Given the Lagrangian in (\ref{action}), the canonically conjugated momentum of the spatial metric has the form
\begin{equation}
\frac{\pi^{ij}}{\sqrt{g}}=G^{ijkl}K_{kl} \,.
\end{equation}
According to the previous discussion about the invertibility of $G^{ijkl}$,
the time derivative of the metric  $\dot{g}_{ij}$ can be completely solved from this expression only if $\lambda\neq1/2$, as we assume.

As happens in the canonical formulation of general relativity \cite{Arnowitt:1962hi}, the Legendre transformation automatically incorporates the momentum constraint 
\begin{equation}
\mathcal{H'}^{i}=-2\nabla_{j}\pi^{ij},
\label{HI'}
\end{equation}
which generates spatial coordinate transformations on the pair  $(g_{ij},\pi^{ij})$. Consequently, the shift function $N_{i}$ can be regarded as the Lagrange multiplier associated to the momentum constraint.  We are interested in the full generator of the spatial diffeomorphisms, therefore we must include the generator of the spatial-coordinate transformations in $(N,P_{N})$ \cite{Donnelly:2011df}, hence we redefine (\ref{HI'}) by
\begin{equation}
\mathcal{H}^{i} = 
-2\nabla_{j}\pi^{ij} + P_{N}\partial^{i}N \,.
\label{Hi}
\end{equation}

Unlike general relativity, the bulk part  of the Hamiltonian does not arise as a sum of the primary constraints, it arises instead as 
\begin{equation}
H=
\int d^{2}x \left[ N \left( \frac{\pi^{ij}\pi_{ij}}{\sqrt{g}} + \frac{\lambda}{1-2\lambda} \frac{\pi^{2}}{\sqrt{g}} + \sqrt{g}\mathcal{V} \right) + N_{i}\mathcal{H}^{i} + \sigma P_{N}\right] \,,
\label{H1} 
\end{equation}
where $\sigma$ is another Lagrange multiplier.

Before we proceed further, we parenthetically make a comment on the definition of asymptotic flatness in the $2+1$ gravitational theories. The exact solution of a particle at rest in $2+1$ general relativity is a flat cone with a deficit angle that depends of the mass of the particle \cite{Deser}. This motivates the definition of asymptotic flatness in $2+1$ general relativity \cite{Ashtekar:1993ds}. In $2+1$ Ho\v rava theory in $2+1$ dimensions  we found \cite{Bellorin:2019zho} that the same solution is valid, hence we postulated the same definition of asymptotic flatness for the three-dimensional Ho\v{r}ava theory. This is stated as the condition of the canonical variables behave asymptotically as \cite{Ashtekar:1993ds} 
 \begin{equation}
 g_{ij}=r^{-\mu}(\delta_{ij}+\mathcal{O}(r^{-1})),\qquad \pi^{ij}\sim \mathcal{O}(r^{\mu-2}), \qquad N=1+\mathcal{O}(r^{-1}),
\label{condiciongpN}
\end{equation}
for any value of the constant $\mu$ (further physical requisites impose some bounds on $\mu$ \cite{Ashtekar:1993ds,Bellorin:2019zho}). The asymptotic flatness condition imposes restrictions on the differentiability of the Hamiltonian (\ref{H1}), hence a counterterm must be added to the Hamiltonian (\ref{H1}),
  \begin{equation}
\mathcal{E}=+2\pi\beta\mu\label{E}\,.
\end{equation}
This energy term is the same as in $2+1$ general relativity, except by the presence of the constant $\beta$.
  
Now we move to the time preservation of the primary constraints. The preservation of the $P_{N} = 0$ generates a secondary constraint, the Hamiltonian constraint $\mathcal{H} = 0$, where
 \begin{eqnarray}
\mathcal{H} &=& 
\frac{1}{\sqrt{g}} \left( \pi^{ij}\pi_{ij} + \frac{\lambda}{1-2\lambda} \pi^{2} \right) 
+ \sqrt{g} \left( \mathcal{V}-\frac{1}{N} B \right) \,,
\label{Hconstr}
\end{eqnarray}
and $B$ stands for total-divergence terms,
\begin{eqnarray}
B &\equiv& 
-2\alpha\nabla_{k}(Na^{k})+4\alpha_{2}\nabla_{k}(Na^{2}a^{k})
+2\alpha_{3}\nabla_{k}(NRa^{k})-\alpha_{4}\left( \nabla^{2}(Na^{2}) \right.
\nonumber \\ &&
\left. -2\nabla_{l}(\nabla_{k} a^{k} N a^{l}) \right)
-\alpha_{5}\nabla^{2}(NR)-2\alpha_{6}\nabla^{kl}(N\nabla_{l}a_{k})-2\alpha_{7}\nabla^{2}(N\nabla_{l}a^{l}) \,.
\nonumber \\ &&
\label{BN}
\end{eqnarray}
Note that the integral of the Hamiltonian constraint (\ref{Hconstr}) has the form
\begin{equation}
\int\,d^{2}xN\mathcal{H}=\int\,d^{2}xN\left(\frac{\pi^{ij}\pi_{ij}}{\sqrt{g}}+\frac{\lambda}{1-2\lambda}\frac{\pi^{2}}{\sqrt{g}}+\sqrt{g}\mathcal{V}\right) -\int\,d^{2}x\sqrt{g}B \,.
\label{HN}
\end{equation}
As a consequence of the asymptotically flat conditions (\ref{condiciongpN}), the last integral in (\ref{HN}), which is a boundary integral, is zero. Contrasting (\ref{HN}) with (\ref{H1}), we see that, at the end, the Hamiltonian (\ref{H1}) can be written as a sum of constraints plus the term (\ref{E}),
\begin{equation}
H=
\int d^{2}x \left( N \mathcal{H} + N_{i}\mathcal{H}^{i} + \sigma P_{N}\right) + 2\pi \beta \mu \,.
\end{equation}

The last constraint that arises under Dirac's procedure is the Hamiltonian constraint (\ref{Hconstr}). Since this analysis involves very large formulas, we show it in the appendix A. We also show in this appendix the canonical evolution equations obtained when all the constraints of the theory are taken into account. Summarizing, we have found two primary constraints, (\ref{Pn}) and (\ref{Hi}), and one secondary constraint (\ref{Hconstr}). 
 
\subsection{Algebra of the constraints}
In this section we show the algebra of the constraints. The Poisson brackets that involve the momentum constraint are
\begin{eqnarray}
\left\{\int\,d^{2}x\epsilon^{k}\mathcal{H}_{k} \,, \int\,d^{2}y\eta^{l}\mathcal{H}_{l}\right\}&=&\int\,d^{2}x\mathcal{H}_{l}\mathcal{L}_{\vec{\epsilon}}\eta^{l},
\\
\left\{\int\,d^{2}x\epsilon^{k}\mathcal{H}_{k} \,, \int\,d^{2}y\eta\mathcal{H}\right\}&=&\int\,d^{2}x\mathcal{H}\mathcal{L}_{\vec{\epsilon}}\eta,
\\
\left\{\int\,d^{2}x\epsilon^{k}\mathcal{H}_{k} \,, \int\,d^{2}y\eta P_{N}\right\}&=&\int\,d^{2}xP_{N}\mathcal{L}_{\vec{\epsilon}}\eta.
\end{eqnarray}
This confirms that the momentum constraint (\ref{Hi}) is a first-class constraint. It is the generator of the gauge symmetry associated to the spatial transformations.

The Poisson bracket of $P_{N}$ with itself is zero. The Hamiltonian constraint (\ref{Hconstr}) does not commute with itself neither with $P_{N}$, therefore the constraints $\mathcal{H}$ and $P_{N}$ are second-class constraints. Indeed, the Poisson brackets between them are
\begin{eqnarray}
&&
\left\{\int\,d^{2}x\epsilon\mathcal{H},\int\,d^{2}y\eta P_{N}\right\} =
\nonumber \\ && 
\int d^{2}x \sqrt{g} \eta' \left[ 
B \epsilon' - 2\nabla^{k}(\epsilon a_{k}f_{2}) 
+ \nabla^{2}(\epsilon f_{3})
+ 2\alpha_{6} \nabla^{kl}(\epsilon\nabla_{l}a_{k})
+ 2Nf_{2}a^{k}\nabla_{k}\epsilon' \right.
\nonumber \\ && 
+ f_{3}N\nabla^{2}\epsilon'
- \nabla^{k}\left(2Nf_{2}\nabla_{k}\epsilon'\right)
- 8\alpha_{2}\nabla_{l}\left(Na^{l}a^{k}\nabla_{k}\epsilon'\right)
+ 2\alpha_{4}\nabla^{2}\left(Na^{k}\nabla_{k}\epsilon'\right)
\nonumber \\ &&  
\left.
-2\alpha_{4}\nabla_{k}\left(Na^{k}\nabla^{2}\epsilon'\right)
+2\alpha_{7}\nabla^{2}\left(N\nabla^{2}\epsilon'\right)
+ 2 \alpha_{6} N \nabla_l a_k \nabla^{kl} \epsilon' 
+ 2 \alpha_6 \nabla_{kl} \left(N\nabla^{kl} \epsilon'\right)
\right],
\nonumber \\ &&
\label{HPN}
\\ &&
\left\{ \int d^{2}x \epsilon \mathcal{H} \,, \int d^{2}y \eta \mathcal{H} \right\} =
\nonumber \\ &&
2 \int d^{2}x \Bigg[
\epsilon \nabla_k \eta \bigg( f_{1}\nabla_{l}\pi^{kl}+\frac{1-\lambda}{1-2\lambda}(\pi\nabla^{k}f_{1}-f_{1}\nabla^{k}\pi)
-\pi^{kl}\nabla_{l}f_{1}-f_{3}\tau^{kl}a_{l}
\nonumber \\ &&
+\tau_{ij}\mathcal{Z}^{ijk}
-\tau a^{k}f_{4} 
+ \tau^{kl} a_{l} ( 2 f_{2} - \alpha_{6}R )
+ 2 (2 \alpha_{2} - \alpha_{4})\tau_{ij}a^{i}a^{j}a^{k}
\nonumber \\ &&
+\frac{\nabla_{l}(Nf_{3})}{N}\left(\frac{1}{2}\tau g^{kl}-\tau^{kl}\right)
+ 2 (\alpha_{4} - 2\alpha_{7})\tau_{ij}\nabla^{i}a^{j}a^{k}
- \nabla^k \Big(\alpha_{4}\tau_{ij}a^{i}a^{j}+\frac{1}{2}\alpha_{5}\tau R
\nonumber \\ &&
+2\alpha_{7}\tau_{ij}\nabla^{i}a^{j}\Big)-\alpha_{4}a^{k}a^{l}\nabla_{l}\tau
+ \frac{\alpha_{6}}{N}\Big( \tau\nabla_{l}(N\nabla^{l}a^{k})
- 2 N\mathcal{M}^{ijk}\tau_{ij}\Big) 
\bigg)
\nonumber \\ &&
+ \eta \left(\tau_{ij}\nabla^{ij} - \tau\nabla^{2} \right) \left( (2\alpha_{3} - \alpha_{6}) \left(Na^{k} \nabla_{k}\epsilon'\right) 
+ \alpha_{5} \left(N \nabla^2 \epsilon'\right) \right)
\nonumber \\ &&
+ 2 \alpha_{4} \eta \tau_{ij} \nabla^{j} \left(N a^i a^l \nabla_l \epsilon' \right)
+ \alpha_7 \eta \left(\tau_{ij}\nabla^{i} - \frac{1}{2} \nabla_{j} \right) \left(Na^{j} \nabla^2 \epsilon' \right)
\nonumber \\ &&
- \alpha_{6} \eta \left(
\tau_{ij} \nabla_{l} \left( N a^{i} a^{j} \nabla^l \epsilon' \right) 
- 2 \tau_{ij} a^j \nabla^{ik} \left( N \nabla_k \epsilon' \right) 
+ \tau \nabla_{l} \left(a^{l} \nabla^{k} \left(N \nabla_k \epsilon' \right) \right) \right)
\nonumber \\ && 
 -(\eta\longleftrightarrow \epsilon) 
\Bigg],
\label{HH}
\end{eqnarray}
where $\epsilon,\eta$ are arbitrary functions and the prime indicates division by $N$, $\epsilon' = \epsilon / N$, $\eta' = \eta / N$. The symbols $f_1$, $f_2$, $f_3$, $f_4$, $\tau^{ij}$, $\tau$, $\mathcal{Z}^{ijk}$ and $\mathcal{M}^{ijk}$ are defined in (\ref{tau}) - (\ref{Mijk}).

The canonical variables are $\{g_{ij},\pi^{ij},N,P_{N}\}$ (8 degrees of freedom). Four functional degrees of freedom are eliminated by the constraints $\mathcal{H}^i$, $P_N$ and $\mathcal{H}$. The gauge symmetry of the spatial diffeomorphisms gives two gauge degrees of freedom. Therefore, among the original eight degrees of freedom in the nonreduced phase space, six are unphysical, leaving two propagating physical degrees of freedom. This represent a even scalar degree of freedom in the theory.

\section{Quantization in the reduced phase space}
\subsection{Linearized theory}
The first scheme we adopt for the quantization of the $2+1$ nonprojectable Ho\v{r}ava theory is to deal with the reduced phase space. To achieve this, we study the linearized theory, since the constraints of the theory can be solved perturbatively. This approach will give us the propagator of the physical mode.

We start with the perturbations around the configuration that corresponds to the ``Minkowski" solution,
\begin{equation}
 g_{ij} = \delta_{ij} \,, 
 \quad
 N = 1 \,,
 \quad
 N_i = 0 \,,
 \quad
 \pi^{ij} = 0 \,.
\end{equation}
The perturbations are parametrized according to
\begin{equation}
g_{ij}=\delta_{ij}+h_{ij}, 
\quad N=1+n,
\quad \pi^{ij}=p^{ij},
\quad N^{i}=n^{i} \,.
\end{equation} 
These linearized variables transform under FDiff as
\begin{eqnarray} 
\delta n&=& \dot{f},
\nonumber\\
\delta n^{i}&=& \dot{\zeta}^{i},\\
\delta h_{ij}&=&2\partial_{(i}\zeta_{j)},
\end{eqnarray}
where the functions $f,\zeta$ are  infinitesimal FDiff parameters. The linear-order version of the field equations (\ref{dotn}) - (\ref{dotpi}) is
\begin{eqnarray}
\dot{n} &=&
\sigma \,,
\label{eqsigma}
\\
\dot{h}_{ij} &=&
2\left(p_{ij}+\frac{\lambda}{1-2\lambda}\delta_{ij}p\right)+2\partial_{(i}n_{j)}
\label{gijl},
\\
\dot{p}^{ij }&=&
\beta \left( \partial^{i}\partial^{j} - \delta^{ij} \partial^{2} \right) n - \left( \partial^{i}\partial^{j} - \delta^{ij} \partial^{2} \right) 
\left( 2\alpha_{1}\left(-\partial^{2}h+\partial_{k}\partial_{l}h^{kl}\right)
+\alpha_{5}\partial^{2}n \right) \,.
\nonumber \\ 
\label{pijl} 
\end{eqnarray}
In appendix A it is shown that the (nonperturbative) solution for $A$ is $A=0$, hence we do not consider this variable in this section. Equation (\ref{eqsigma}) is just an equation for fixing $\sigma$.

We introduce the orthogonal transverse and longitudinal decomposition
\begin{equation}
h_{ij}=\left(\delta_{ij}-\frac{\partial_{i}\partial_{j}}{\partial^{2}}\right)h^{T}+\partial_{(i}h_{j)},
\label{Transvlong}
\end{equation}
and similarly for $p^{ij}$. We impose  the transverse gauge 
\begin{equation}
\partial_{i}h_{ij}=0,
\end{equation}
under which all the longitudinal sector of the metric is eliminated, $h_{i} = 0$. The linearized momentum constraint (\ref{Hi}) eliminates the longitudinal sector of $p^{ij}$, since it takes the form
\begin{equation}
\partial_{i}p^{ij} = 0,
\end{equation}
hence $p_{i}=0$. The  Hamiltonian constraint at first order is given by
\begin{eqnarray}
\beta\partial^{2}h^{T}-\alpha_{5}\partial^{4}h^{T}
+2\alpha\partial^{2}n
+ 2 ( \alpha_{6} + \alpha_{7}) \partial^{4}n &=& 0 \,.
\label{linearhamconst}
\end{eqnarray}
We may solve it for $n$,
\begin{eqnarray}
n&=&\frac{1}{2}\left(\frac{-\beta+\alpha_{5}\partial^{2}}{\alpha+(\alpha_{6}+\alpha_{7})\partial^{2}}\right)h^{T}\nonumber\\
&\equiv& \mathcal{P}h^{T}\label{npt}.
\end{eqnarray}
Therefore, the momentum and Hamiltonian constraints and the transverse  gauge fix the variables $h_{i}$, $p_{i}$ and $n$, leaving the transverse sector $\{h^{T},p^{T}\}$ and the Lagrange multiplier $n_{i}$ active.
The longitudinal sector of the evolution equation (\ref{gijl}) yields an equation for $n_{i}$,
\begin{eqnarray}
\partial^{2}n_{i}+\partial^{j}\partial_{i}n_{j}&=&-\frac{2\lambda}{1-2\lambda}\partial_{i}p^{T},
\end{eqnarray}
whose solution is
\begin{equation}
n_{i}=-\frac{\lambda}{1-2\lambda}\partial_{i}\left(\frac{1}{\partial^{2}}p^{T}\right).
\label{nisolution}
\end{equation}
The longitudinal sector of Eq.~(\ref{pijl}) yields no new information. The traces of the linearized Eqs.~(\ref{gijl}) and (\ref{pijl}) lead automatically to their transverse sector. The trace of these equations, after using eq. (\ref{nisolution}), leads to 
\begin{eqnarray}
\dot{h}^{T} &=& 
2\left(\frac{1-\lambda}{1-2\lambda}\right)p^{T},\\
\dot{p}^{T} &=& -\beta\partial^{2} n -2\alpha_{1}\partial^{4}h^{T}
+\alpha_{5}\partial^{4}n \,.
\end{eqnarray}
These equations, after substituting (\ref{npt}), imply
\begin{eqnarray}
\ddot{h}^{T} &=&
2\left(\frac{1-\lambda}{1-2\lambda}\right)
(-\beta\mathcal{P}\partial^{2}h^{T}
+[\alpha_{5}\mathcal{P} - 2\alpha_{1}]\partial^{4}h^{T}) \,.
\end{eqnarray}
This represent the propagating equation for the scalar mode  $\{h^{T},p^{T}\}$ of the complete nonprojectable Ho\v rava theory in $2+1$ dimensions \cite{Sotiriou:2011dr}. 


\subsection{The reduced Hamiltonian and the propagator of the physical mode }
The physical or reduced Hamiltonian of the linearized theory is obtained by expanding the Hamiltonian (\ref{H1}) up to second order in perturbations (constraints are no added since in the reduced theory they are explicitly solved). The expansion yields
\begin{equation}
\begin{split}
H &= \int\,d^{2}x\Big( p^{ij}p_{ij}+\frac{\lambda}{1-2\lambda} p^{2}-\beta\Big(-\partial^{2}h+\partial_{i}\partial_{j}h^{ij}+h^{ij}\partial^{2}h_{ij}
-\frac{1}{2} h^{ij}\partial_{i}\partial_{l}h^{l}{}_{j}\\
&
+\frac{1}{2} h^{ij} \partial_{i}\partial_{j}h
-\frac{1}{4}\partial_{l}h^{ij}\partial^{l}h_{ij}-\partial_{l}h^{lk}\partial_{i}h^{i}{}_{k}+\frac{1}{2}\partial_{l}h^{lk}\partial_{k}h
-\frac{1}{4} \partial_k h \partial_k h
\\
&
+\left(n+\frac{1}{2}h\right)\left(-\partial^{2}h+\partial_{i}\partial_{j}h^{ij}\right)\Big) -\alpha\partial_{k}n\partial^{k}n +\alpha_{1}\left(-\partial^{2}h+\partial_{i}\partial_{j}h^{ij}\right)^{2}\\
&
+\alpha_{5}\partial^{2}n\left(-\partial^{2}h+\partial_{i}\partial_{j}h^{ij}\right)
+\alpha_{6}\partial^{i}\partial^{j}n\partial_{i}\partial_{j}n+\alpha_{7}\partial^{2}n\partial^{2}n\Big) \,.
\end{split}\label{Hlinear}
\end{equation}
After substituting the solutions for the unphysical variables found in the previous section (in the transverse gauge), we arrive at the reduced Hamiltonian for the transverse sector,
\begin{equation}
\begin{split}
H_{RED}&=
\int\,d^{2}x\left(p^{T} M p^{T} +h^{T} \mathbb{M} h^{T}\right),
\label{HRED}
\end{split}
\end{equation}
where 
\begin{eqnarray}
M &=& \frac{1-\lambda}{1-2\lambda} \,,\\
\mathbb{M} &=&
\left(\beta\mathcal{P}+\alpha\mathcal{P}^{2}\right)\partial^{2}+\left(\alpha_{1}
-\alpha_{5}\mathcal{P}+[\alpha_{6}+\alpha_{7}]\mathcal{P}^{2}\right)\partial^{4} \,.
\end{eqnarray}

We look for the propagator of the independent physical mode in the transverse gauge. The path integral in the  reduced phase space has the form
\begin{equation}
\mathcal{Z}_{0}=\int\,\mathcal{D}h^{T}\,\mathcal{D}p^{T} 
\exp \left[i\int\,dt\,d^{2}x \left(p^{T}\dot{h}^{T}-\mathcal{H}_{RED}\right)\right] \,,
\end{equation}
where $\mathcal{H}_{RED}$ is  reduced Hamiltonian density of the equation (\ref{HRED}). After Gausssian integration in $p^{T}$ we obtain the path integral in noncanonical form
\begin{equation}
\mathcal{Z}_{0}=
\int\,\mathcal{D}h^{T} \exp\left[i\int\,dt\,d^{2}x 
\left(\frac{1}{4 M} \dot{h}^{T}\dot{h}^{T}
- h^T \mathbb{M} h^T \right)\right] \,,
\end{equation}
we can get the propagator of the physical mode in Fourier space,
\begin{equation}
<h^T h^T>=
\frac{1}{\omega^{2}/4 M -\left(\beta\mathcal{P}+\alpha\mathcal{P}^{2}\right)\vec{k}^{2}
+\left(\alpha_{1} - \alpha_{5}\mathcal{P}
+[\alpha_{6}+\alpha_{7}]\mathcal{P}^{2}\right)\vec{k}^{4}} \,.
\end{equation}

\section{The path integral in the nonreduced phase space}
\subsection{Nonperturbative formalism}
\subsubsection{Definition of the measure}
In the Hamiltonian formulation of a field theory, the recipe for the measure of the gauge sector was provided by Faddeev \cite{Faddeev:1969su}. In this formulation, the gauge symmetries have first-class constraints associated. The measure is then given by the brackets between the first-class constraints and the chosen gauge-fixing conditions, which, by definition, must be nonzero brackets. In our case the first class constraints are the components of the momentum constraints $\mathcal{H}^i$. Let us denote by $\chi^i = 0$ the associated gauge-fixing conditions. The measure of the gauge sector is given by
\begin{equation}
 \det\{\mathcal{H}^{k},\chi^{l}\} \,.
\label{firstclassmeasure}
\end{equation}

The path-integral quantization of systems with second-class constraints was focused by Senjanovic in Ref.~\cite{Senjanovic:1976br}. He showed that the measure associated to the second-class constraints is 
\begin{equation}
 \sqrt{\det\{\theta_{p},\theta_{q}\}} \,,
 \label{measuredef}
\end{equation}
where $\theta_p$ stands for each one of the second-class constraints. Note that this definition only makes sense in the canonical formulation, where Poisson brackets are defined. This is one of the reasons why we focused the Hamiltonian formulation of the $2+1$ nonprojectable Ho\v{r}ava theory. Senjanovic's proof of his theorem starts with postulating the path integral (for a mechanical system) with the measure (\ref{measuredef}), then the integration over the unphysical variables leads to a canonical path integral over the reduced phase space with measure 1, as it must be. An essential step of the proof is the existence of a canonical transformation that leads to transformed second-class constraints with ``canonical" Poisson brackets between them, which means that these brackets are equal to a given constant matrix. After this transformation, the integration on unphysical variables can be done, leading to the reduced path integral with measure 1. Fradkin and Fradkina \cite{Fradkin:1977xi} introduced the same measure (\ref{measuredef}) for systems with second-class constraints, extending the so-called Batalin-Fradkin-Vilkovisky quantization of gauge theories (in \cite{Fradkin:1977xi} fermionic degrees of freedom were also included). An alternative proof of the measure (\ref{measuredef}), based on a geometrical approach, can be found in the book of Henneaux and Teitelboim \cite{Henneaux:1992ig}.

In order to use the measure (\ref{measuredef}) in the present theory, we introduce a common notation for the second-class constraints, namely, $\theta_{1}=\mathcal{H}$ and $\theta_{2}=P_{N}$. The path integral in terms of the nonreduced canonical variables has then the form
\begin{equation}
Z_{0}=
\int \mathcal{D}V \delta(\mathcal{H}^{i}) \delta(\chi^{i}) \delta(\theta_1) \delta(\theta_2) e^{iS_{can}},
\end{equation}
where the measure and the action are given respectively by
\begin{equation}
\mathcal{D}V = 
\mathcal{D} g_{ij} \mathcal{D}\pi^{ij} \mathcal{D}N \mathcal{D}P_{N} \times \det\{\mathcal{H}^{k},\chi^{l}\} \sqrt{\det\{\theta_{p},\theta_{q}\}} \,,
\label{medida}
\end{equation}
\begin{equation}
S_{can}=
\int\,dt\,d^{2}x \left( P_N \dot{N} + \pi^{ij}\dot{g_{ij}} -\frac{N}{\sqrt{g}}\left(\pi^{ij}\pi_{ij}+\frac{\lambda}{1-2\lambda}\pi^{2}\right)-N\sqrt{g}\mathcal{V}\right)\label{scan} \,.
\end{equation}

\subsubsection{The measure of the gauge sector}
Since we have the momentum constraint explicitly, Eq.~(\ref{Hi}), we can make explicit computations on the measure of the gauge sector, taking a quite general gauge-fixing condition $\chi^i$. This must be a condition that fixes the freedom of choosing spatial coordinates. For simplicity, let us consider that this condition only involves the spatial metric, $\chi^i = \chi^i(h)$, but otherwise arbitrary. Using the expression (\ref{Hi}), we obtain the bracket
\begin{equation}
\{\mathcal{H}^{k}(x),\chi_{l}(y)\}=2\int\,d^{2}z\nabla_{(i}\left(\delta^{k}_{j)}\delta^{2}\right)\dfrac{\delta\chi_{l}(h)}{\delta h_{ij}}=-2\delta^{k}_{(i}\nabla_{j)}\left(\dfrac{\delta\chi_{l}}{\delta h_{ij}}\right).\label{hichi}
\end{equation}

A important question one can pose here is whether this approach for dealing with the gauge fixing and the first-class constraints is equivalent to the usual approach (Faddeev-Popov) for incorporating the gauge-fixing condition in the path integral quantization of gauge theories, typically formulated in terms of covariant Lagrangians. To answer this question, we evaluate the Fadeev-Popov determinant of the  standard approach. We need the gauge-transformed field, where the gauge symmetry is an infinitesimal spatial diffeomorphism, then
\begin{equation}
 h^\zeta_{ij} = h_{ij} + 2\nabla_{(i}\zeta_{j)} \,.
\end{equation}
The Faddeev-Popov factor becomes
\begin{equation}
\dfrac{\delta\chi_{l}(h^\zeta (x))}{\delta\zeta_{k}(y)} = \int\,d^{2}z\dfrac{\delta\chi_{l}(h^{\zeta}(x))}{\delta h_{ij}^{\zeta}(z)}\dfrac{\delta h_{ij}^{\zeta}(z)}{\delta\zeta_{k}(y)}=
2 \int d^{2}z \nabla_{(i}\left( \delta^{k}_{j)}\delta^{2}\right) \dfrac{\delta\chi_{l} (h^{\zeta})}{\delta h^{\zeta}_{ij}} \,.
\end{equation}
This shows, at least for a general gauge-fixing condition that depends on the spatial metric, $\chi^i(h)$, that the measure in the Hamiltonian formulation (\ref{hichi}) and the measure in the Fadeev-Popov approach coincide.

If we consider the transverse gauge condition, $\chi_j = \partial_i h_{ij} = 0$, the bracket of the measure becomes
\begin{equation}
\{\mathcal{H}^{k},\chi_{l}\} =\dfrac{\delta\chi_{l}(h')}{\delta\zeta_{k}}=\delta^{k}_{l}\partial^{2}\delta^{2}+\partial^{k}\partial_{l}\delta^{2}+\left(\delta^{k}_{l}\Gamma_{ij}{}^{i}+\Gamma_{jl}{}^{k}\right)\partial^{j}\delta^{2}.
\end{equation}
In the path integral, the determinant of the gauge sector is incorporated to the action by means of ghosts fields,
\begin{equation}
\det\{\mathcal{H}^{k},\chi_{l}\}
= 
\int \mathcal{D}\bar{\epsilon} \, \mathcal{D}\epsilon \exp\Big(i\int dt d^{2}x \, \bar{\epsilon}_{k} \Big( \delta^{k}_{l}\partial^{2}\delta^{2} + \partial^{k}\partial_{l}\delta^{2} + \left(\delta^{k}_{l}\Gamma_{ij}{}^{i}+\Gamma_{jl}{}^{k}\right)\partial^{j}\delta^{2}\Big)\epsilon^{l}\Big) \,.
\end{equation}


\subsubsection{The second-class-constraint measure}
There is a important simplification because the matrix of brackets between the second-class constraints acquires a triangular form,
\begin{equation}
\{\theta_{p},\theta_{q}\}=\left(\begin{array}{cc}
\{\mathcal{H},\mathcal{H}\} & \{\mathcal{H},P_{N}\}\\
\{P_{N},\mathcal{H}\}& 0
\end{array}\right) \,,
\end{equation}
which is a consequence of the fact that $P_N$ has zero bracket with itself. Then the determinant of the matrix of Poisson brackets is a quadratic form, hence
\begin{equation}
\sqrt{\det\{\theta_{p},\theta_{q}\}}=
\det\{\mathcal{H},P_{N}\} \,,
\label{sprtbra}
\end{equation}
simplifying greatly the measure of the second-class constraints. The bracket between the Hamiltonian constraint and $P_{N}$ is show in (\ref{HPN}). By setting $\epsilon$ and $\eta$ as appropriate Dirac deltas in (\ref{HPN}), we obtain
\begin{eqnarray}
&&
\{\mathcal{H}(\vec{x})\,,P_{N}(\vec{y})\}=
\nonumber \\ &&
\frac{\sqrt{g}}{N} \left[ 
\frac{B}{N}
- 2 \nabla^{k}(. a_{k}f_{2})
+ \nabla^{2}(. f_{3})
+ 2 \alpha_{6}\nabla^{kl} (.\nabla_{l}a_{k})
+ 2 N f_{2}a^{k}\nabla_{k}\left(\frac{.}{N}\right)
\right.
\nonumber \\ &&
+f_{3}N\nabla^{2}\left(\frac{.}{N}\right)
- 2 \nabla^{k}\left(Nf_{2}\nabla_{k}\left(\frac{.}{N}\right)\right)
-8\alpha_{2}\nabla_{l}\left(Na^{l}a^{k}\nabla_{k}\left(\frac{.}{N}\right)\right)
\nonumber \\ &&
+2\alpha_{4}\nabla^{2}\left(Na^{k}\nabla_{k}\left(\frac{.}{N}\right)\right)
-2\alpha_{4}\nabla_{k}\left(Na^{k}\nabla^{2}\left(\frac{.}{N}\right)\right)
+2\alpha_{7}\nabla^{2}\left(N\nabla^{2}\left(\frac{.}{N}\right)\right)
\nonumber \\ &&
\left.
+ 2 \alpha_{6} N \nabla_{l}a_{k} \nabla^{kl} \left(\frac{.}{N}\right)
+ 2 \alpha_{6} \nabla_{kl} \left( N \nabla^{kl} \left(\frac{.}{N}\right)\right) \right] \delta^{(2)}(\vec{x}-\vec{y}) \,.
\end{eqnarray}
(the dot means $\nabla_i (a \cdot ) b = \nabla_i (a b)$). We may promote the measure (\ref{sprtbra}) to the action by means of ghosts fields. We define the fermionic fields $\eta$ and $\bar{\eta}$, such that
\begin{equation}
\det\{\mathcal{H},P_{N}\}=
\int\mathcal{D}\bar{\eta}\,\mathcal{D}\eta
\exp\left(i\int dt d^2x \, \bar{\eta}\,\{\mathcal{H},P_{N}\}\,\eta\right) \,.
\label{gostsecond}
\end{equation}

\subsection{Perturbations in the path integral: linearized theory}
We the aim of focusing the ultraviolet behavior of the theory, in this section we consider only the $z=2$ terms and discard the lower order terms in the Lagrangian. Thus, in this section we will find the most important terms for the propagators of the nonreduced variables at the ultraviolet regime.

According to the definition of the measure (\ref{medida}), the result (\ref{sprtbra}), and expanding the action up to second order in perturbations, the perturbative path integral takes the form
\begin{eqnarray}
\mathcal{Z}_0 &=& 
\int \mathcal{D} h_{ij} \mathcal{D} p_{ij} \mathcal{D} n \mathcal{D} p_n \times
\delta(\mathcal{H}^i) \delta(\chi_j) \delta(\mathcal{H}) \delta(p_n)
\det\{\mathcal{H}^{k},\chi^{l}\} \det\{\mathcal{H},p_n\}
\nonumber
\\ &&
\exp \Bigg[
i \int dt d^{2}x
\Bigg( p_n \dot{n} + p^{ij}\dot{h}_{ij} 
-\left( p^{ij}p_{ij}+\frac{\lambda}{1-2\lambda}p^{2} \right)
-\alpha_{6}\partial^{i}\partial^{j}n\partial_{i}\partial_{j}n 
\nonumber
\\ &&
-\alpha_{7}\partial^{2}n\partial^{2}n
-\alpha_{1}\left(-\partial^{2}h+\partial_{i}\partial_{j}h^{ij}\right)^{2}
-\alpha_{5}\left(-\partial^{2}h+\partial_{i}\partial_{j}h^{ij}\right)\partial^{2}n\Bigg) \Bigg] \,.
\nonumber
\\
\label{scanper}
\end{eqnarray}
The measure corresponding to the second-class constraints can be promoted to the Lagrangian by means of ghost fields. The ghosts are of first order in perturbations, therefore in the quadratic action the determinant of second-class constraints must be considered at zero order,
\begin{equation}
\det\{\mathcal{H},p_n\}
=
\int\mathcal{D}\bar{\eta}\,\mathcal{D}\eta
\exp\left(i\int dt d^{2}x \,\bar{\eta}\left(2(\alpha_{6}+\alpha_{7})\partial^{4}\right)\eta\right) \,.
\label{measurezero}
\end{equation}

To further advance in the computations, and since we are mainly interested in the role of the second-class-constraint sector, we perform a partial reduction in the phase space eliminating all the variables that belong to the gauge sector, but leaving the second-class constraints unsolved. This allows us to find the propagators of all the sector that is not associated to the gauge symmetry. First, let us decompose the spatial metric $h_{ij}$ and its conjugate momentum $p_{ij}$ in the transverse+longitudinal decomposition shown in (\ref{Transvlong}),
\begin{eqnarray}
S &=& 
 \int dt d^{2}x \Bigg[ 
p_n \dot{n} 
+ M \left (p^{T} - \frac{1}{2} \left(\frac{1}{M}\dot{h}^{T}
-\frac{2\lambda}{1-\lambda} \partial_{k}p_{k} \right)\right)^{2} 
\nonumber
\\ &&
+\frac{1}{2}(p_{i}-\dot{h}_{i}) 
\left(\delta^{i}_{k}\partial^{2} + \partial^{i}\partial_{k}\right) p^{k}
+\frac{1}{4}M \left( \frac{1}{M}\dot{h}^{T}-\frac{2\lambda}{1-\lambda}\partial^{k}p_{k} \right)^{2}
\nonumber \\ &&
+(M-1)p_{k}\partial^{k}\partial^{l}p_{l}
-(\alpha_{6}+\alpha_{7})n\partial^{4}n
-\alpha_{1}(h^{T}+\partial^{k}h_{k})\partial^{4}(h^{T}+\partial^{k}h_{k})
\nonumber \\ &&
+\alpha_{5}n\partial^{4}(h^{T}+\partial^{k}h_{k}) \Bigg]
\end{eqnarray}
To eliminate the gauge sector, we impose the transverse gauge, $\chi_{j}=\partial_i h_{ij} = 0$, which eliminates $h_i$. Moreover, the linearized momentum constraint eliminates $p_i$ upon integration. The measure of the gauge sector $\det\{\mathcal{H}^{k},\chi^{l}\}$ yields an irrelevant (constant) factor at second order in perturbations. The resulting path integral is 
\begin{eqnarray}
&&
\mathcal{Z}_{0}  = 
\nonumber \\ &&
\int\mathcal{D} h^{T} \mathcal{D} p^{T}\mathcal{D}n\mathcal{D}p_{N} \mathcal{D}\bar{\eta} \mathcal{D}\eta \delta(\mathcal{H})\delta(p_{N})
\exp \Bigg[ i\int dt d^{2}x\Bigg( p_{N}\dot{n}
-M\Big(p^{T}-\frac{1}{2M}\dot{h}^{T}\Big)^{2}
\nonumber \\ &&
+\frac{1}{4 M} \dot{h}^{T}\dot{h}^{T}
-\Big( \alpha_ 1 h^{T} \Delta^{2} h^{T}+ n((\alpha_{6}+\alpha_{7})\Delta^{2})n-h^{T}(\alpha_{5}\Delta^{2})n\Big)
\nonumber \\ &&
+\bar{\eta}\left(2(\alpha_{6}+\alpha_{7})\partial^{4}\right)\eta  \Bigg)\Bigg] \,,
\end{eqnarray}
The variables $p_n$ and $p^T$ are not associated to gauge symmetries, but them can be integrated directly ($p_n$ is trivial and the integration on $p^T$ is Gaussian). Finally, we may promote the delta $\delta(\mathcal{H})$ to the Lagrangian by means of a Lagrange multiplier, which we denote by $a$, 
\begin{equation}
\delta(\mathcal{H})=
\int \mathcal{D}a \exp\Big(i\int dt d^{2}x \, a \mathcal{H}\Big) \,,
\end{equation}
such that the path integral includes the integration in $a$. In the perturbative approach, $a$ is considered as a variable of linear order, hence, in the quadratic action we need the expression of the Hamiltonian constraint $\mathcal{H}$ up to linear order. The linear-order expression for $\mathcal{H}$ can be taken from Eq.~(\ref{linearhamconst}),
\begin{eqnarray}
\mathcal{H} = 
-\alpha_{5} \partial^{4}h^{T} 
+ 2 ( \alpha_{6} + \alpha_{7}) \partial^{4} n \,.
\end{eqnarray}
After these steps, we arrive at the path integral
\begin{eqnarray}
\mathcal{Z}_{0} &=&
\int \mathcal{D} h^{T} \mathcal{D} n \mathcal{D} \bar{\eta}\, \mathcal{D}\eta 
\exp \Bigg[ 
i \int dt d^{2}x\Bigg(
\frac{1}{4 M} \dot{h}^{T}\dot{h}^{T} 
- \alpha_{1} h^{T} \partial^4 h^{T} 
- (\alpha_{6}+\alpha_{7}) n \partial^4 n
\nonumber \\ &&
+ \alpha_{5} h^{T} \partial^4 n 
+ a \left( -\alpha_{5} \partial^{4}h^{T} 
+ 2 ( \alpha_{6} + \alpha_{7}) \partial^{4} n \right)
+ 2 (\alpha_{6} + \alpha_{7}) \bar{\eta} \partial^{4} \eta  \Bigg)\Bigg] \,.
\label{quadraticlagrangian}
\end{eqnarray}
From this quantum action we may extract the propagators of the $h^T$ field and the variables associated to the second-class constraints, which are $n$, $a$, $\bar{\eta}$, $\eta$. Our notation for the propagators is as follows: in Fourier space, the quadratic Lagrangian given in (\ref{quadraticlagrangian}) is written as
\begin{equation}
\left( \begin{array}{ccccc}
h^{T} & n & a & \bar{\eta} & \eta 
\end{array}\right) 
\left(\begin{array}{cc}
\mathfrak{M}_{1} & 0 \\
0 & \mathfrak{M}_{2} \\
\end{array}\right)
\left(\begin{array}{c}
h^{T} \\
n \\
a \\
\bar{\eta}\\ 
\eta
\end{array}\right) \,,
\label{matrix}
\end{equation}
where 
\begin{eqnarray}
&& 
\mathfrak{M}_{1}=
\left(\begin{array}{ccc}
\frac{\omega^{2}}{4 M} + \alpha_{1}k^{4} 
& -\alpha_{5}k^{4}/2
& \alpha_5 k^4/2
\\
-\alpha_{5}k^{4}/2 
& (\alpha_{6}+\alpha_{7})k^{4} 
& -(\alpha_6 + \alpha_7) k^4
\\
\alpha_5 k^4/2
& -(\alpha_6 + \alpha_7) k^4
& 0
\end{array}\right) \,,
\\ &&
\mathfrak{M}_{2}=
\left(\begin{array}{cc}
0 & - ( \alpha_{6} + \alpha_{7} ) k^{4}
\\
(\alpha_{6}+\alpha_{7} ) k^{4} & 0 
\end{array}\right) \,.
\end{eqnarray}
Therefore, to get the propagators we must invert the matrix in (\ref{matrix}). This yields the propagators
\begin{eqnarray}
<h^{T}h^{T}> &=&
4 M  \left(\omega^{2}+ M \left(4\alpha_{1}-\frac{\alpha_{5}^{2}}{\alpha_{6}+\alpha_{7}}\right)k^{4}\right)^{-1},
\\
<nn> &=&
\frac{\alpha_{5}^{2} M}{(\alpha_{6}+\alpha_{7})^{2}}\left(\omega^{2} + M \left(4\alpha_{1}-\frac{\alpha_{5}^{2}}{\alpha_{6}+\alpha_{7}}\right)k^{4}\right)^{-1}  \,,
 \label{propagatorn}
\\
<h^{T}n>
&=& 
\frac{2\alpha_{5} M} {\alpha_{6}+\alpha_{7}}\left(\omega^{2} + M \left(4\alpha_{1} - \frac{\alpha_{5}^{2}}{\alpha_{6}+\alpha_{7}}\right)k^{4}\right)^{-1},
\\
< h^T a > &=&
0 \,,
\\
< n a > &=&
-\frac{1}{(\alpha_6 + \alpha_7) k^4} \,,
\\
< a a > &=&
-\frac{1}{(\alpha_6 + \alpha_7) k^4} \,,
\\
<\bar{\eta} \eta > &=& 
\frac{1}{(\alpha_6 + \alpha_7) k^4}.
\label{ghostsprop}
\end{eqnarray}
We observe that the propagators of $h^T$ and $n$ get a regular form. Any nonregular part of the propagator of the lapse function has disappeared once the second-class constraint have been taken into account. However, the propagators of the auxiliary variables $a$ and $\bar{\eta},\eta$ and the mixed correlator between $n$ and $a$ acquire nonregular forms. There is an evident similarity between these nonregular propagators.

The propagator of the ghosts $\bar{\eta},\eta$ (\ref{ghostsprop}) is restricted by the fact that these fields are of first order in perturbations. Due to this, in the linearized theory the measure of the second-class constraints contributes only at zero order in perturbations, in the sense we commented before (\ref{measurezero}). This implies that, in the resulting linearized quantum Lagrangian, there are not couplings between the ghosts $\bar{\eta},\eta$ and the rest of the fields. Note that this criterium of order in perturbations is independent of the gauge chosen; in any gauge only the zero-order terms of the second-class-constraint measure contribute, as it is exemplified in (\ref{measurezero}) in the transverse gauge. There are two consequences of this restriction that have incidence on the form of the propagator (\ref{ghostsprop}). The first one is that the matrix in (\ref{matrix}) has always a block-diagonal form, with no crossings between the $\bar{\eta},\eta$ sector and the other sectors. The propagator of $\bar{\eta},\eta$ is then the direct inverse of the block $\mathfrak{M}_2$; it does not receive contributions from the other fields. In this way, the dependence on $\omega$ arising in $\mathfrak{M}_1$ due to other fields cannot be transferred to the propagator of the $\bar{\eta},\eta$ ghosts. The second consequence concerns the possibility of generating $\omega$-dependence on the $\bar{\eta},\eta$ propagator directly from the measure (\ref{gostsecond}), by using another gauge-fixing condition (that is, getting dependence on $\omega$ directly in $\mathfrak{M}_2$). This would require that the measure (\ref{gostsecond}) depends on some canonical momentum, such that time derivatives are generated after integration in momentum. But this is also discarded by the order criterium: the coupling between the $\bar{\eta},\eta$ ghosts and any canonical momentum would be, at least, of cubic order in perturbations. In conclusion, although the form of the propagator (\ref{ghostsprop}) may depend on the gauge chosen, it cannot acquire dependence on $\omega$, hence it can not be brought to a regular form.

In Ref.~\cite{Barvinsky:2015kil} the renormalization of the projectable Ho\v{r}ava theory was achieved. Those authors obtained regular propagators by imposing a gauge-fixing condition that in particular is nonlocal. The projectable theory has no second-class constraints. In the quantization we have presented for the nonprojectable theory, the freedom to impose different gauge-fixing conditions is limited to canonical gauges, which are gauge conditions that depend only on the canonical variables. The transverse gauge is an example of a canonical gauge. This is so because Faddeev's measure of the gauge sector (\ref{firstclassmeasure}) is defined only for canonical gauge conditions $\chi_i$. We are forced to take this measure for the gauge sector because we define the path integral in the canonical formalism, which in turn is a requisite imposed by the presence of second-class constraints. Therefore, the gauge-fixing condition used in Ref.~\cite{Barvinsky:2015kil} cannot be inserted here since it is a noncanonical gauge condition. It depends on the shift vector $N_i$, which is a Lagrange multiplier from the point of view of the canonical formalism.

Finally, we comment that nonlocal but canonical gauges do not render the propagator of the $\bar{\eta},\eta$ regular due to the same previous reason, since nonlocality of operators does not alter the criterium of order in perturbations discussed above.

Although the linearized theory does not allow to explore further, when considering amplitudes with interactions there is still the possibility that some cancellations occur between the propagators of the ghosts fields $\bar{\eta},\eta$ and the rest of fields. Aimed as a first step towards this task, in appendix B we present the theory at the first nontrivial order for interactions: the cubic theory. 


\section{Conclusions}
We have considered the consistent quantization of the $2+1$ nonprojectable Ho\v{r}ava theory, considering all the terms in the Lagrangian that are covariant under the FDiff gauge symmetry. Our central focus has been the presence of the second-class constraints, which requires that the quantization must be addressed in a different way to the pure gauge theories, that is, gauge theories without second-class constraints. We highlight that the recipe for the measure of a theory with second-class constraints is known in the Hamiltonian formalism \cite{Senjanovic:1976br}, see also \cite{Fradkin:1977xi,Henneaux:1992ig}. In this paper we have evaluated this formula for the measure explicitly for the 2+1 Ho\v{r}ava theory.

We have performed the full Hamiltonian analysis of the theory, finding all the constraints explicitly and classifying them between first and second class. As expected, the momentum constraint is the only first-class constraint. It is associated to the symmetry of arbitrary spatial diffeomorphisms over each leaf of the foliation, which is the only gauge symmetry of the theory in the strict sense. The set of constraints is complete in the sense that its preservation leads to elliptic differential equations for Lagrange multipliers. An application of this analysis is the characterization of the physical propagating mode of the theory. We have shown this by means of a perturbative analysis.

As we commented, the Hamiltonian formulation of the theory has enabled us to obtain the measure of the second-class constraints explicitly. A central result is that the matrix of Poisson brackets of the second-class constraints acquires a quadratic form. This simplifies the square root of the measure, such that it can be incorporated to the quantum Lagrangian by means of fermionic ghosts. We have extracted further consequences of the measure using perturbative analysis. We have found that, when the Hamiltonian constraint and the measure are considered in the quantization procedure, the propagator of the lapse function acquires a completely regular form. There are no nonregular terms in this propagator, nor in the one of the scalar mode of the spatial metric. However, auxiliary variables associated to the second-class constraints (Lagrange multiplier and ghosts) acquire nonregular propagators. Nonregular terms were previously found in Ref.~\cite{Barvinsky:2015kil}, presenting an obstacle to achieve the renormalization of the nonprojectable theory by means of the technique of regular propagators. We have discussed how the nonregular form of the propagator of the ghosts associated the second-class constraints can not be altered (it cannot acquire a dependence on the frequency) by choosing a different canonical gauge-fixing condition (throughout this paper we have used the transverse gauge). In the quantization of the projectable Ho\v{r}ava theory \cite{Barvinsky:2015kil}, the choosing of the appropriated gauge condition, in particular a nonlocal gauge, leaded to regular propagators. Thus, the nonlocal gauge condition was crucial for obtaining the renormalization of the projectable theory. We have discussed that such a gauge condition cannot be imposed in this scheme of quantization since it is a noncanonical gauge. A further possibility is the usage of a canonical but nonlocal gauge-fixing condition, but this does not alter the nonregularity of the propagator of the ghosts associated to the second-class constraints.

Further computations on the quantization of the theory with interactions are required. In particular they may shed light on the renormalization, since explicit amplitudes can be computed. An interesting question is whether there are cancellations between the auxiliary variables associated to the second-class constraints.  With this aim, we have presented the quantum theory, that is, the measure and the potential, at cubic order in perturbations. We expect that this work can be extended with the computations of the (scalar) graviton scattering and to explore its renormalization.


\section*{Acknowledgement}

B.~D.~is partially supported by the CONICYT PFCHA/DOCTORADO BECAS CHILE/2019 - 21190398, and by the grant PROYECTO ANT1856 of Universidad de Antofagasta, Chile.

\appendix

\section{Consistency of the set of constraints}
In section 3.1 we arrived at the Hamiltonian constraint  (\ref{Hconstr}), which arises as a secondary constraint. In this appendix we continue on Dirac's procedure for obtaining the constraints of the theory. The next step is to impose the preservation of the Hamiltonian constraint (\ref{Hconstr}). This is obtained from the bracket
\begin{eqnarray}
&&
	\{H,\int\,d^{2}x\eta\mathcal{H}\}  = 
\nonumber \\ &&
	\int d^{2}x \Bigg\{ \mathcal{H}\mathcal{L}_{\vec{N}}\eta
	+ 2 (N\nabla_{k}\eta - \eta\nabla_{k}N) \Bigg[f_{1}\nabla_{l}\pi^{kl}+(1+\omega)(\pi\nabla^{k}f_{1}-f_{1}\nabla^{k}\pi) 
\nonumber \\ && 
	-\pi^{kl}\nabla_{l}f_{1} -f_{3}\tau^{kl}a_{l}
	+\tau_{ij}\mathcal{Z}^{ijk}
	- \tau a^{k}f_{4}
	+ \tau^{kl}a_{l} ( 2 f_{2}- \alpha_{6}R) 
	+2 (2\alpha_{2}- \alpha_{4})\tau_{ij}a^{i}a^{j}a^{k} 
\nonumber \\ &&
	+\frac{\nabla_{l}(Nf_{3})}{N}\Big(\frac{1}{2}\tau g^{kl}-\tau^{kl}\Big)
	+ 2 (\alpha_{4}- 2\alpha_{7})\tau_{ij}\nabla^{i}a^{j}a^{k} -\alpha_{4}a^{k}a^{l}\nabla_{l}\tau
	-\nabla^{k}\Big(\alpha_{4}\tau_{ij}a^{i}a^{j}
\nonumber \\ &&
	+\frac{\alpha_{5}}{2}\tau R+2\alpha_{7}\tau_{ij}\nabla^{i}a^{j}\Big)
	+ \frac{\alpha_{6}}{N} \left( \tau \nabla_{l}(N\nabla^{l}a^{k})
	- 2 N\mathcal{M}^{ijk}\tau_{ij} \right)\Bigg]
\nonumber \\ &&
	+ 2N \Bigg[ (-2\alpha_{3}+\alpha_{6}) \left(\tau_{ij} \nabla^{ij} -\tau\nabla^{2} \right) \left( N a^{k} \nabla_k \eta' \right)
	- \alpha_{5} \left( \tau_{ij} \nabla^{ij} -\tau\nabla^{2} \right) \left(N \nabla^2 \eta'\right) 
\nonumber \\ &&	
	-2 \alpha_{4} \tau_{ij} \nabla^{j} \left( N a^i a^l  \nabla_l \eta' \right)
	+ \alpha_{6} \left( \tau_{ij} \nabla_{l} \left( N a^i a^j \nabla^l \eta' \right) 
	- 2 \tau_{ij} a^{j} \nabla^{ik} ( N \nabla_{k}\eta' ) \right.
\nonumber \\ &&
    \left.
	- \tau \nabla_{l} \left( a^{l} \nabla^k (N \nabla_k \eta')\right) \right) 
	- \alpha_{7} \left( \tau_{ij} \nabla^{i} \left( N a^j \nabla^2 \eta' \right) 
	- \frac{1}{2} \nabla_l \left( N a^l \nabla^2 \eta'\right) \right) \Bigg]
\nonumber \\ && 
	-\sqrt{g}\frac{\sigma}{N} \bigg[ 
	B \eta'
	- 2 \nabla^{k}(\eta a_{k}f_{2})
	+ \nabla^{2}(\eta f_{3}) 
	+ 2 \alpha_{6} \nabla^{kl} (\eta\nabla_{l}a_{k})
    + 2 Nf_{2}a^{k}\nabla_{k}\eta'
\nonumber \\ &&	
	+f_{3}N\nabla^{2}\eta'
	- 2 \nabla^{k}\left( N f_{2}\nabla_{k}\eta'\right)
	- 8 \alpha_{2}\nabla_{l}\left(Na^{l}a^{k}\nabla_{k}\eta'\right)
	+ 2 \alpha_{4}\nabla^{2}\left(Na^{k}\nabla_{k}\eta'\right)
\nonumber \\ &&
	- 2 \alpha_{4}\nabla_{k}\left(Na^{k}\nabla^{2}\eta'\right)
	+ 2 \alpha_{7}\nabla^{2}\left(N\nabla^{2}\eta'\right)
	+ 2 \alpha_{6} N \nabla_{l}a_{k} \nabla^{kl} \eta'
    + 2 \alpha_{6} \nabla_{kl} \left( N \nabla^{kl} \eta'\right)
    \bigg] \Bigg\}  \,,
\nonumber \\ &&	
\label{HnH}
\end{eqnarray}
where $\eta$ is arbitrary function and the prime indicates division by $N$, $\eta'={\eta}/{N}$. The symbols introduced above are
\begin{eqnarray}
\tau_{kl}&=&\pi_{kl}+\frac{\lambda}{1-2\lambda} g_{kl}\pi, \qquad\tau=g^{kl}\tau_{kl},
\label{tau}
\\
f_{1} &=&
-\beta+2\alpha_{1}R+\alpha_{3}a^{2}+\alpha_{5}\nabla_{k}a^{k}, \\ 
f_{2} &=&
-\alpha +\alpha_{3}R+\alpha_{4}\nabla_{k}a^{k}+2\alpha_{2}a^{2}, \\ 
f_{3} &=&\alpha_{4}a^{2}+\alpha_{5}R+2\alpha_{7}\nabla_{k}a^{k},\\ 
f_{4} &=&
-\alpha+2\alpha_{2}a^{2}+\alpha_{4}\left(a^{2}+\nabla_{l}a^{l}\right)
+ \alpha_{5} R + \frac{1}{2}\alpha_{6} R,\\
\mathcal{Z}^{(ij)k} &=&
\frac{1}{2}f_{3}g^{ij}a^{k}-\alpha_{6}\left( a^{(i}\nabla^{k}a^{j)}+a^{(i}\nabla^{j)}a^{k}-a^{k}\nabla^{(i}a^{j)}\right),\\
\mathcal{M}^{ijk}&=&a^{i}\nabla^{j}a^{k}+a^{l}\nabla_{l}a^{i}g^{kj}+g^{ki}\nabla^{j}\nabla_{l}a^{l}.
\label{Mijk}
\end{eqnarray}
To obtain the desired time preservation of the Hamiltonian constraint, we put $\eta=\delta^{2}(x-y)$ into the bracket (\ref{HnH}). This yields an elliptic partial differential equation for the Langrange multiplier $\sigma$,
\begin{eqnarray}
&&
	0=
	- \sqrt{g} \Bigg[ 
	2 ( \alpha_{6} + \alpha_{7} ) 
	\left( \frac{1}{N} \nabla^4 + 2 a^k \nabla_k \nabla^2 \right) \sigma'
	+ 2 \left(f_{3}-f_{2}-\frac{\alpha_{4}}{N} \nabla_{k}\left(Na^{k}\right)
	\right.
\nonumber \\ &&
    \left.
	+\frac{2 \alpha_{7}}{N} \nabla^{2}N \right) \nabla^{2}\sigma'
	+\Big(-8\alpha_{2}a_{l}a_{k} + \frac{4 \alpha_{4}}{N} \nabla_{k}\left(Na_{l}\right)+\alpha_{6}\Big(4\nabla_{l}a_{k}
	+\frac{2}{N}\nabla_{kl} N
\nonumber \\ &&
	\Big)\Big)\nabla^{kl} \sigma'
	+\frac{1}{N} \Big( 2 \nabla_{k}\left(N(f_{3}-f_{2})\right)
	- 8 \alpha_{2} \nabla_{l}(Na^{l}a_{k})
	+ 2 \alpha_{4} \nabla^{2} \left(Na_{k}\right)
\nonumber \\ &&
	+ 4 \alpha_{6} \nabla^{l} \left(N\nabla_{k}a_{l}\right)\Big)\nabla^{k}\sigma'
	+ \frac{1}{N} \Big( NB
	- 2 \nabla_{k}\left(Nf_{2}a^{k}\right)
	+ 2 \alpha_{6}  \nabla^{kl} \left(N\nabla_l a_k\right)
\nonumber \\ &&
	+ \nabla^{2}\left(Nf_{3}\right)\Big)\sigma' \Bigg]
	+ 2 (2\alpha_{3}-\alpha_{6}) \nabla_{k}\left(Na^{k}
	\left(\nabla^{ij} (N\tau_{ij})
	-\nabla^{2}\left(N\tau\right)\right)\right)
\nonumber \\ &&
	-2\alpha_{5} \nabla^2 \left( N
	\left( \nabla^{ij} (N\tau_{ij})-\nabla^{2}\left(N\tau\right) \right) \right)
	-4\alpha_{4}\nabla_{k}\Big(Na^{k}a^{i}\nabla^{j}(N\tau_{ij})\Big)
\nonumber \\ &&
	+ 2 \alpha_{6} \nabla^{l} \left( N a^{i}a^{j} \nabla_{l} \left(N\tau_{ij}\right)
	+ 2 N \nabla^{i} \nabla_{l}\left(Na^{i}\tau_{ij}\right)
	- N\nabla_{l}\left(\tau a^{k} \nabla_{k} (N\tau)\right) \right)
\nonumber \\ &&
	+ \alpha_{7} \nabla^2 \left( 2 Na^{i} \nabla^{j}(N\tau_{ij}) - N^{2} a_k a^k \right)
	- 2 \nabla_{k}(N P^k) - 2 N a_{k} P^k -\nabla_{k}\left(\mathcal{H}N^{k}\right)
	\,,
\nonumber \\
\label{Eqsigma'}
\end{eqnarray}
where 
\begin{eqnarray}
	&&
	P^k =
	f_{1}\nabla_{l}\pi^{kl}+(1+\omega)(\pi\nabla^{k}f_{1}-f_{1}\nabla^{k}\pi)-\pi^{kl}\nabla_{l}f_{1} -f_{3}\tau^{kl}a_{l}
	+\tau_{ij}\mathcal{Z}^{ijk}
\nonumber \\ &&
	-\tau a^{k} f_{4} + \tau^{kl} a_{l} (2 f_{2} - \alpha_{6} R) 
	+ 2 (2 \alpha_{2}- \alpha_{4})\tau_{ij}a^{i}a^{j}a^{k}
	+\frac{\nabla_{l}(Nf_{3})}{N}\left(\frac{1}{2}\tau g^{kl}-\tau^{kl}\right)
\nonumber \\ &&
	+ 2 ( \alpha_{4}- 2 \alpha_{7})\tau_{ij}\nabla^{i}a^{j}a^{k}
	- \nabla^{k}(\alpha_{4}\tau_{ij}a^{i}a^{j}+\frac{\alpha_{5}}{2}\tau R+2\alpha_{7}\tau_{ij}\nabla^{i}a^{j})
	-\alpha_{4}a^{k}a^{l}\nabla_{l}\tau
\nonumber \\ &&   
	+ \frac{\alpha_{6}}{N}\left( \tau\nabla_{l}(N\nabla^{l}a^{k}) 
	- N\mathcal{M}^{ijk}\tau_{ij}\right) \,.
\end{eqnarray}
The preservation of the Hamiltonian constraint $\mathcal{H}$ has generated the equation (\ref{Eqsigma'}) for the Lagrange multiplier $\sigma$. This is an elliptic partial differential equation, whenever the condition $\alpha_6 + \alpha_7 \neq 0$ on the space of coupling constants holds (this is the coefficient of the highest order (elliptic) operator on $\sigma$). Therefore, this equation can be solved with the proper boundary conditions. Dirac's procedure ends at this step, no more constraints are generated.

 The Hamiltonian with all the constraints incorporated takes the form 
 \begin{equation} 
 H=
 \int d^{2}x \left[ N \left( \frac{\pi^{ij}\pi_{ij}}{\sqrt{g}} + \frac{\lambda}{1-2\lambda} \frac{\pi^{2}}{\sqrt{g}} + \sqrt{g}\mathcal{V} \right) 
 	+ N_{i}\mathcal{H}^{i}+\sigma P_{N}+A\mathcal{H}\right]
 	+ 2\pi \beta \mu ,\label{H1A}
 \end{equation} 
 where $A$ is a Lagrange multiplier (we assume that $A$ decays fast enough asymptotically). The variations of the canonical action with respect to the canonical variables $(g_{ij},\pi^{ij})$ and $(N,P_N)$ yields the following field equations
 \begin{eqnarray}
 &&
 \dot{N} = 
 N^{k}\nabla_{k}N + \sigma \,, 
 \label{dotn}
 \\ &&
 \dot{g}_{ij} = \frac{2}{\sqrt{g}}\left(N+A\right)\left(\pi_{ij}+\frac{\lambda}{1-2\lambda} g_{ij}\pi\right)+2\nabla_{(i}N_{j)} \,,
\\ &&
 \frac{\dot{\pi}^{ij}}{\sqrt{g}} =
\nonumber \\ &&
 	(N+A)\Bigg(
 	\frac{1}{2}\frac{g^{ij}}{g}\left(\pi^{kl}\pi_{kl}+\frac{\lambda}{1-2\lambda}\pi^{2}\right)
 	-\frac{2}{g}\left(\pi^{ik}\pi^{j}{}_{k}+\frac{\lambda}{1-2\lambda}\pi\pi^{ij}\right)
 	-\frac{1}{2}g^{ij}\mathcal{V}
\nonumber \\ && 
  	+\left(
  	\frac{1}{2}f_{1}g^{ij}R 
  	+f_{2}a^{i}a^{j}
  	+f_{3}\nabla^{i}a^{j}
  	+2\alpha_{6}\nabla^{i}a^{k}\nabla^{j}a_{k}\right)\Bigg)
  	-(\nabla^{ij} (f_{1}(N+A))
\nonumber \\ &&
    -g^{ij}\nabla^{2}(f_{1}(N+A))+\nabla^{(i}(f_{3}a^{j)}(N+A))
    -\nabla_{k}(\mathcal{Z}^{(ij)k}(N+A))) 
    -\mathcal{H}^{(i}N^{j)}
\nonumber \\ &&
 	- 2\nabla_{k}(N^{(i}\pi^{kj)})+\nabla_{k}(N^{k}\pi^{ij})
 	+ \alpha N ( a^{k}g^{ij} - 2 a^{(i}g^{j)k} ) \nabla_k A'
 	- 2 \alpha_{2} N \left( a^{k}g^{ij} \right.
\nonumber \\ && \left.
    - 2 a^{k}a^{i}a^{j} - 2 a^{2}a^{(i}g^{kj)}\right) \nabla_{k}A'
 	+ 2 \alpha_{3} \left( N Ra^{(i}g^{kj)} \nabla_{k}A'
 	- \nabla^{ij}\left(N a^{k} \nabla_{k}A'\right) \right.
\nonumber \\ &&
    \left.
	+ g^{ij}\nabla^{2}\left(N a^{k} \nabla_{k}A'\right) \right)
 	+\alpha_{4} \bigg( \frac{1}{2} \left( g^{kl}g^{ij} -2g^{k(i}g^{j)l} \right) \nabla_{l} (N a_m a^m ) \nabla_{k} A'+Na^{i}a^{j} \nabla^2 A'
\nonumber \\ && 
 	-\left(\nabla_{l}a^{l}a^{k} -2\nabla_{l}a^{l}a^{(i}g^{kj)} -2a^{k}\nabla^{i}a^{j}\right)N \nabla_{k}A'
 	-2\nabla^{(i}\left(a^{j)}N a^{k}\nabla_{k}A'\right)
\nonumber \\ &&
 	+g^{ij}\nabla_{k}\left(a^{k}N a^{l}\nabla_{l} A'\right) \bigg)
 	+ \frac{\alpha_{5}}{2} \left( \left( g^{kl}g^{ij} -g^{k(i}g^{j)l} \right) \nabla_{kl}A'
 	+ g^{ij} N R \nabla^2 A' \right.
\nonumber \\ &&
    \left.
 	- 2 \nabla^{ij}\left(N \nabla^2 A'\right)
 	+ 2 g^{ij}\left(N\nabla^2 A'\right) \right)
 	+ \alpha_{6}\Big( g^{ij}\nabla_{l}\left(N\nabla^{l}a^{k}\right) \nabla_{k}A'
 	+ \nabla^k \left(Na^{i}a^{j} \nabla_k A' \right)
\nonumber \\ &&
 	- 2 \left( \nabla^{(ij)}a^{k} + \nabla^{l}N \nabla_{l}a^{(i}g^{j)k} \right) \nabla_{k}A'
 	- \frac{1}{2} g^{ij} N R a^{k} \nabla_k A'
	+ \nabla^{ij}\left(Na^{k} \nabla_k A'\right)
\nonumber \\ &&
 	- N \nabla^{(i}A' \left( 2 \nabla^{j)} \nabla_{l}a^{l} + Ra^{j)} \right)
	- g^{ij}\nabla^{2}\left(Na^{k} \nabla_k A'\right)
	+ 2 \nabla^i a^j \nabla^k \left(N \nabla_k A'\right)
\nonumber \\ &&
 	- 2 \nabla^{(i}\left(a^{j)} \nabla^k\left(N \nabla_k A'\right)\right)
    + g^{ij} \nabla_l \left(a^l \nabla^k \left(N \nabla_k A'\right) \right)\Big)
    + \alpha_7 \left(  2 N\nabla^{i}a^{j}\nabla^2 A' 
    \right.
\nonumber \\ &&
    \left.
 	+ \left( g^{kl}g^{ij} -2 g^{k(i}g^{j)l} \right) \nabla_l (N\nabla_m a^m) \nabla_k A'
    - 2 \nabla^{(i}\left(a^{j)}N'\nabla^2 A'\right)
 	+ g^{ij}\nabla_{l}\left(Na^{l}\nabla^2 A'\right) \right) \,,
\nonumber \\
\label{dotpi} 
\\ &&
0 = 
\mathcal{H} - \nabla^{i}(P_{N}N_{i}) + \frac{\sqrt{g}}{N} \Big( BA' 
- 2 \nabla^{k} (A a_{k}f_{2})
+ \nabla^{2}(A f_{3})
+ 2 \alpha_{6} \nabla^{kl} (A\nabla_{l}a_{k})
\nonumber \\ && 
+ 2 Nf_{2}a^{k}\nabla_{k}A'
+ f_{3}N\nabla^{2}A'
- 2 \nabla^{k}\left( N f_{2}\nabla_{k}A'\right)
- 8 \alpha_{2} \nabla_{l} \left(Na^{l}a^{k}\nabla_{k}A'\right)
\nonumber \\ &&
+ 2\alpha_{4}\nabla^{2}\left(Na^{k}\nabla_{k}A'\right)
- 2\alpha_{4}\nabla_{k}\left(Na^{k}\nabla^{2}A'\right)
+ 2\alpha_{7}\nabla^{2}\left(N\nabla^{2}A'\right)
\nonumber \\ &&
+ 2\alpha_{6}N \nabla_{l}a_{k} \nabla^{kl}A'
+ 2\alpha_{6}\nabla_{kl} \left(N\nabla^{kl} A'\right)\Big)	\,.
\label{eqA}
\end{eqnarray} 
Equation (\ref{eqA}) corresponds to the variations with respect to $N$. Since $P_N = 0$ is a preserved constraint of the theory, this equation acquires no time derivative. Instead, it is an elliptic partial differential equation for the Lagrange multiplier $A$. It is of fourth-order, $\alpha_6 + \alpha_7$ being the coefficient of the fourth-order operator. Hence we require that this combination is not zero. The same condition arose previously when preserving the constraints. Since we assume that $A$ goes asymptotically (fast enough) to zero, the only everywhere-continuous solution is $A= 0$.

\section{Interactions}
The analysis we have presented can be continued by incorporating interactions. This is a highly nontrivial step due to the fact that the number of interactions (vertices) is big. As a contribution to this task, in this appendix we list the third-order version of the measure of the second-class constraints and the potential
\begin{eqnarray}
&&
\det\{\mathcal{H},P_{N}\} =
\nonumber \\ && 
\int \mathcal{D}\bar{\eta}\,\mathcal{D}\eta
\exp\bigg[ 
i\int dt d^{2}x\, \bar{\eta} \Big(-2(\alpha_{6}+\alpha_{7})\Big(-\partial^{4}+\partial^{4}n+n\partial^{4}
+2\partial_{k}n\partial^{2}\partial^{k}\Big)
\nonumber \\ &&
+(4\alpha_{4}-2\alpha_{6}-8\alpha_{7})\partial_{i}\partial_{j}n\partial^{i}\partial^{j}+(-3\alpha_{4}-2\alpha_{6}
+2\alpha_{7})\partial^{2}n\partial^{2}
\nonumber \\ &&
+ 2(-\alpha_{3}+\alpha_{5})\left(-\partial_{k}\partial^{2}h+\partial_{k}\partial_{i}\partial_{j}h^{ij}\right)\partial^{k}
+(-\alpha_{3}+2\alpha_{5})\left(-\partial^{2}h+\partial_{i}\partial_{j}h^{ij}\right)\partial^{2}
\nonumber \\ &&
+(\alpha_{6}+\alpha_{7})\Big(h\partial^{4}
-2\partial^{2}\left(\delta^{ij}\chi_{ijk}\partial^{k}\right)\Big)+\alpha_{6}\Big(-\partial_{k}\partial^{2}h\partial^{k}+\partial_{k}\partial_{i}\partial_{j}h^{ij}\partial^{k}
-\partial^{2}h\partial^{2}
\nonumber \\ &&
+\partial_{i}\partial_{j}h^{ij}\partial^{2}-2\delta^{ij}\chi_{ijk}\partial^{k}\partial^{2}\Big)+2\alpha_{7}\Big(-\partial^{2}h^{ij}\partial_{i}\partial_{j}
-2\partial^{i}h^{jk}\partial_{i}\partial_{j}\partial_{k}-2h^{ij}\partial^{2}\partial_{i}\partial_{j}
\nonumber \\ &&
-\partial_{i}h^{ij}\partial_{j}\partial^{2}+\frac{1}{2}\partial_{k}h\partial^{k}\partial^{2}\Big)\Big)\eta \bigg] \,,
\end{eqnarray}
where
\begin{equation}
\chi_{ijk} =
\frac{1}{2} 
\left( \partial_i h_{jk} + \partial_j h_{ik} - \partial_k h_{ij} \right) \,.
\end{equation}

The potential up to third order takes the form
\begin{eqnarray}
&&	
N\sqrt{g}\mathcal{V} =
\nonumber \\ &&
\alpha_{6}\partial^{i}\partial^{j}n\partial_{i}\partial_{j}n+\alpha_{7}\partial^{2}n\partial^{2}n+\alpha_{1}\left(-\partial^{2}h+\partial_{i}\partial_{j}h^{ij}\right)^{2}
+\alpha_{5}\left(-\partial^{2}h+\partial_{i}\partial_{j}h^{ij}\right)\partial^{2}n
\nonumber \\ &&
+\alpha_{4}\partial^{2}n\partial^{k}n\partial_{k}n-\alpha_{6}\Big(2\partial_{k}n\partial^{i}n\partial^{k}\partial_{i}n
+n\partial_{k}\partial^{i}n\partial^{k}\partial_{i}n\Big)
\nonumber \\ &&
-\alpha_{7}\left(2\partial^{k}n\partial_{k}n\partial^{2}n+n\partial^{2}n\partial^{2}n\right)+\alpha_{1}n\left(-\partial^{2}h+\partial_{i}\partial_{j}h^{ij}\right)^{2}
\nonumber \\ && 
+\alpha_{3}\left(-\partial^{2}h+\partial_{i}\partial_{j}h^{ij}\right)\partial^{k}n\partial_{k}n+\alpha_{5}\Bigg(\left(-\partial^{2}h+\partial_{i}\partial_{j}h^{ij}\right)\Bigg(\frac{1}{2}h\partial^{2}n
-h^{kl}\partial_{k}\partial_{l}n-\partial^{k}n
\nonumber \\ &&
\partial_{k}n-\partial^{k}n\gamma^{lp}\chi_{lpk}\Bigg)+\Bigg(h^{ij}\partial^{2}h_{ij}-2h^{ij}\partial_{i}\partial_{l}h^{l}{}_{j}
+h^{ij}\partial_{i}\partial_{j}h+ \frac{1}{2}\partial_{l}h^{ij}\partial^{l}h_{ij}
\nonumber \\ &&
-\partial_{l}h^{lk}\partial_{i}h^{i}{}_{k}
+\frac{1}{2}\partial_{l}h^{lk}\partial_{k}h\Bigg)\partial^{2}n\Bigg)
+\alpha_{6}\Bigg(-2h^{ij}\partial^{k}\partial_{j}n\partial_{k}\partial_{i}n+\frac{1}{2}h\partial^{i}\partial^{j}n\partial_{i}\partial_{j}n
\nonumber \\ &&
-2\chi_{ijk}\partial^{i}\partial^{j}n\partial^{k}n\Bigg)
+\alpha_{7}\Bigg(\frac{1}{2}h\partial^{2}n\partial^{2}n-2h^{ij}\partial_{i}\partial_{j}n\partial^{2}n-2\partial^{2}n\partial^{i}n\delta^{jk}\chi_{jki}\Bigg)
\nonumber \\ &&
+\alpha_{1}\Bigg(\frac{1}{2}h\left(-\partial^{2}h+\partial_{i}\partial_{j}h^{ij}\right)^{2}
+2\left(-\partial^{2}h+\partial_{k}\partial_{l}h^{kl}\right)\Big(h^{ij}\partial^{2}h_{ij}
-2h^{ij}\partial_{i}\partial_{r}h^{r}{}_{j}
\nonumber \\ &&
+h^{ij}\partial_{i}\partial_{j}h+\frac{1}{2}\partial_{r}h^{ij}\partial^{r}h_{ij}-\partial_{r}h^{rj}\partial_{i}h^{i}{}_{j}+\frac{1}{2}\partial_{i}h^{ij}\partial_{j}h\Big)\Bigg) \,.
\end{eqnarray}
The Hamiltonian constraint up to third order takes the form 
\begin{eqnarray}
&&
	\mathcal{H} = p^{ij}p_{ij}+\omega p^{2}-\beta\Big(R_{(1)}+R_{(2)}+\frac{1}{2}hR_{(1)}\Big)-\alpha\partial_{k}n\partial^{k}n+\alpha_{1}R_{(1)}^{2}
	+\alpha_{5}\partial^{2}nR_{(1)}
\nonumber \\ &&
	\alpha_{6}\partial^{i}\partial^{j}n\partial_{i}\partial_{j}n+\alpha_{7}\partial^{2}n\partial^{2}n+
	2\alpha\left([1-n]\partial^{2}n+\partial_{k}\left(\frac{1}{2}h\partial^{k}n-h^{kl}\partial_{l}n\right)\right)
\nonumber \\ &&
	-2\alpha_{3}\partial_{k}( R_{(1)}\partial^{k}n)
	+\alpha_{4}\partial^{2}(\partial_{l}n\partial^{l}n)-2\alpha_{4}\partial_{l}(\partial^{l}n\partial^{2}n)+\alpha_{5}\Big([1-n]\partial^{2}R_{(1)}
\nonumber \\ &&
	-\partial_{k}\Big((h^{kl}-\frac{1}{2}\delta^{kl}h)\partial_{l}R_{(1)}\Big)+\partial_{2}(nR_{(1)}+R_{(2)})\Big)+
	2\alpha_{6}\Bigg([\delta^{kl}-h^{kl}-n\delta^{kl}]\partial_{l}(\partial^{2}\partial_{k}n)
\nonumber \\ &&
	+\frac{1}{2}\partial^{k}(h\partial^{2}\partial_{k}n)+\partial^{k}\Big(
	-\frac{1}{2}h^{ij}\partial_i\partial_{j}\partial_{k}n-\partial_{2}(n\partial_{k}n)-\partial^{i}(\partial_{l}n\gamma^{lm}\chi_{ikm})+\partial^{i}(n\partial_{i}\partial_{k}n)
\nonumber \\ &&
	-\delta^{ij}\partial^{m}\partial_{k}n\chi_{ijm}-\partial^{j}\partial_{l}n\gamma^{lm}\chi_{kjm}\Big)\Bigg)+2\alpha_{7}\Bigg((1-n)(\partial^{4}n)+
	\partial_{k}\Big((-h^{kl}+\frac{1}{2}h\delta^{kl})\partial_{l}\partial^{2}n\Big)
\nonumber \\ &&
	+\partial^{2}\Big(n\partial^{2}n-\partial^{j}(n\partial_{j}n)-\delta^{ij}\delta^{lm}\partial_{l}n\chi_{ijm}
	-h^{ij}\partial_{i}\partial_{j}n\Big)\Bigg) \,.
\end{eqnarray}

where 
\begin{eqnarray}
R_{(1)}&=&
-\partial^{2}h+\partial_{i}\partial_{j}h_{ij},
\\ 
R_{(2)}&=&
[n+\frac{1}{2}h](-\partial^{2}h+\partial_{i}\partial_{j}h^{ij})+\frac{1}{2}h^{ij}\partial_{i}\partial_{j}h
-\frac{1}{2}h^{ij}\partial^{k}(\partial_{i}h_{jk}+\partial_{j}h_{ik}-\partial_{k}h_{ij})
\nonumber \\ &&
-\partial_{k}(h^{kl}\partial^{i}h_{il})+\frac{1}{2}\partial_{k}(h^{kl}\partial_{l}h)+\frac{1}{2}\partial^{k}(h^{ij}\partial_{k}h_{ij})
+\frac{1}{2}\partial^{k}h\partial^{l}h_{kl}
-\frac{1}{4}\partial_{k}h\partial^{k}h
\nonumber \\ &&
-\frac{1}{4}(-\partial_{i}h_{jk}\partial^{i}h^{jk}+2\partial_{i}h_{jk}\partial^{k}h^{ij})
\end{eqnarray}


\end{document}